\documentclass[aps,pre,groupedaddress]{revtex4-2} 
\usepackage{natbib}\label{key}
\usepackage{amsmath}
\usepackage{epsfig}
\usepackage{color}

\usepackage{graphicx}
\usepackage{amsfonts}
\usepackage{dcolumn}
\usepackage{bm}
\usepackage{color}
\usepackage{hyperref}

\begin{document}
\bibliographystyle{unsrt}

\title{Asymptotic Large Deviations of Counting Statistics in Open Quantum Systems}
\author{Fei Liu}
\email[Email address: ]{feiliu@buaa.edu.cn}
\affiliation{School of Physics, Beihang University, Beijing 100191, China}

\date{\today}

\begin{abstract}
We use a semi-Markov process method to calculate large deviations of counting statistics for three open quantum systems, including a resonant two-level system and resonant three-level systems in the $\Lambda$- and $V$-configurations. In the first two systems, radical solutions to the scaled cumulant generating functions are obtained. Although this is impossible in the third system, since a general sixth-degree polynomial equation is present, we still obtain asymptotically large deviations of the complex system. Our results show that, in these open quantum systems, the large deviation rate functions at zero current are equal to two times the largest nonzero real parts of the eigenvalues of operator $-{\rm  i}\hat H$, where $\hat H$ is a non-Hermitian Hamiltonian, while at a large current, these functions possess a unified formula.   
\end{abstract}
\maketitle

\section{Introduction} 
\label{section1}
In the past two decades, the counting statistics of open quantum systems~\cite{Levitov1996,Bagrets2003,Esposito2009} have attracted considerable theoretical interest~\cite{Bruderer2014,Garrahan2010,Carollo2019,Esposito2008,Zinidarifmmodeheckclseci2014,Gasparinetti2014,Buca2014,Cuetara2015,Zheng2003,Hasegawa2020,Brandes2008,Budini2011,Rudge2019}. A widely accepted method is the tilted quantum master equation (TQME)~\cite{Mollow1975,Zheng2003,Bagrets2003,Esposito2009,Liu2016a}. 
For instance,  to obtain the key scaled cumulant generating functions (SCGFs) of large deviations~\cite{Touchette2008}, one can diagonalize the generator of the TQME and solve for its largest real eigenvalue~\cite{Bagrets2003,Esposito2009}. The TQME is very similar in form to the quantum master equation (QME). Assume that the dimension of a quantum system is $D$. Then, the generator is a $D^2\times D^2$ matrix. Because the size of the matrix rapidly increases with the  dimension, the large deviation properties of open quantum systems are usually investigated numerically. There are few analytical results~\cite{Garrahan2010,Budini2011,Zinidarifmmodeheckclseci2014,Buca2014,Liu2022}

\begin{figure}
\includegraphics[width=1.\columnwidth]{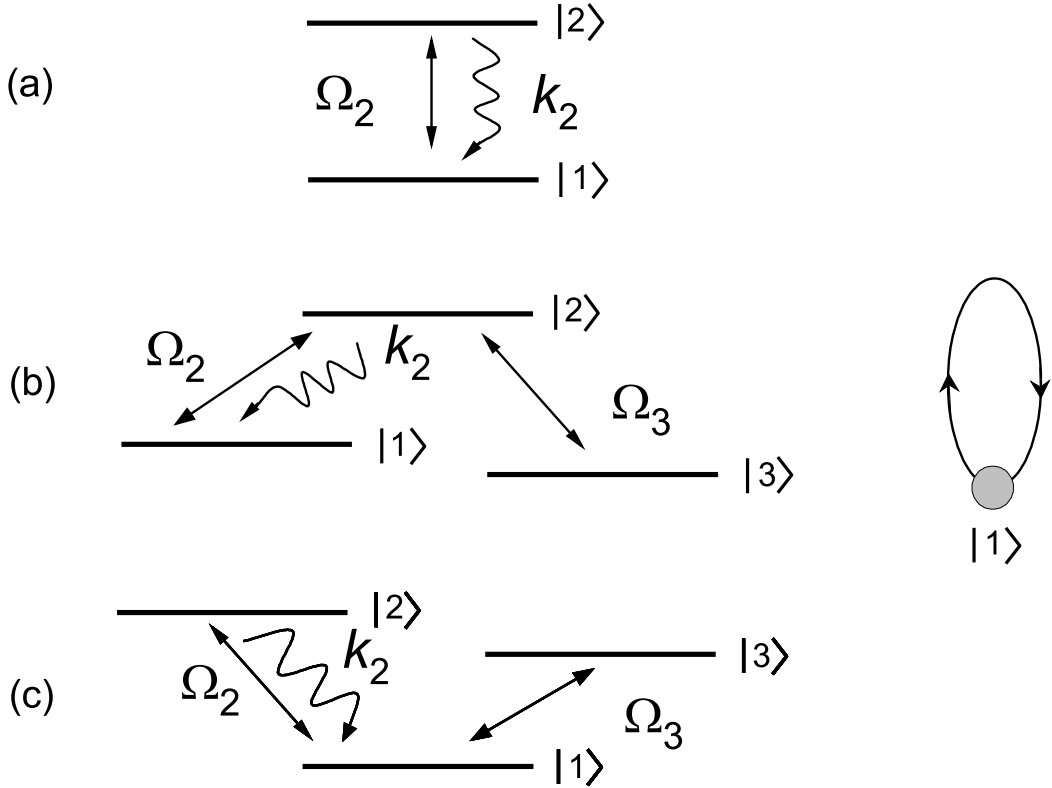}
\caption{Schematic diagrams of 
one two-level and two three-level quantum systems. In the left column, the lines with double arrows indicate resonant driving, and the lines with a single arrow represent spontaneous decays. In the right column, the gray dot  represents the states $|1\rangle$ of the quantum systems. The curve denotes the deterministic evolution of the wave functions that depart from the states. The arrows on the curve point to the state $|1\rangle$ to which each wave function eventually collapses. }
\label{fig1}
\end{figure}

Very recently, we proposed a semi-Markov process (sMP) method to study this topic~\cite{Liu2022}. This method is directly based on the probability interpretation of quantum jump trajectories ~\cite{Mollow1975,Dalibard1992,Carmichael1993,Plenio1998,Breuer2002,Wiseman2010}. These trajectories, which pertain to the evolution of the wave functions of single quantum systems, are composed of deterministic pieces and random collapses of the wave functions. Figure~(\ref{fig1}) illustrates this picture: the gray dot in the right column represents the states $|1\rangle$ of the three quantum systems (a)-(c) from which the quantum trajectories depart, the curve denotes the deterministic evolution of the wave functions, and the arrows indicate the state to which the wave functions collapse. Because the time intervals between departure and termination are distributed nonexponentially and are independent of previous histories, these processes are  sMPs~\cite{Ross1995,Qian2006,Andrieux2008,Carollo2019,Liu2022}. In contrast to the TQME method, the dimension in the sMPs method is equal to the number of collapsed states, $M$, and the size of the involved generator is $M\times M$~\cite{Andrieux2008,Esposito2008,Liu2022}. $M$ is usually smaller than $D$. Hence, it is interesting to explore whether this method can reveal some unnoticed large deviation properties of quantum systems in an analytical way. In this work, we report several findings for resonant two-level systems and three-level quantum systems in $\Lambda$- and $V$-configurations. These open quantum systems are typical and have wide applications in quantum optics~\cite{Mandel1995,Scully1997} and quantum thermodynamics~\cite{Alicki2018}.  

This paper is organized as follows. In Sec.~(\ref{section2}), we exactly solve the SCGF of a simple two-level system and analyze its asymptotic behaviors. A general computing scheme is developed. Section~(\ref{section3}) studies a three-level quantum system in the $\Lambda$-configuration. We find that the SCGF of the system is indeed identical to that of the two-level system. Section~(\ref{section4}) focuses on asymptotically large deviations of a more challenging three-level system in the $V$-configuration. To achieve this aim, we determine the parameter space of this nontrivial quantum system. In Sec.~(\ref{section5}), we unify these asymptotic SCGFs and derive corresponding large deviation rate functions. Section~(\ref{section6}) concludes this paper.

\section{Two-level system}
\label{section2}
We start with the simple quantum system shown in Fig.~(\ref{fig1})(a). It is driven by a resonant field, and the surrounding environment is a vacuum. The QME generator of the system is
\begin{eqnarray}
\label{QMEtwolevel}
{\cal L}[\rho]=-{\rm i}\left[ H,\rho\right] + k_{2}[\sigma_{12}\rho\sigma_{21} -\frac{1}{2}\{\sigma_{21}\sigma_{12}, \rho \} ] .
\end{eqnarray}
Here, the Planck constant $\hbar$ is set to 1, 
\begin{eqnarray}
	H=-\frac{\Omega_{2}}{2}(\sigma_{12} +\sigma_{21})
\end{eqnarray} 
is the interaction Hamiltonian between the system and the field in the interaction picture, $\sigma_{12}\equiv |1\rangle \langle 2|$, $\sigma_{21}\equiv |2\rangle \langle 1|$,  $\Omega_{2}$ is the Rabi frequency of the field driving the $|1\rangle$-$|2\rangle$ transition, and $k_{2}$ is the decay rate. The two-level system is assumed to depart from the ground state $|1\rangle$ at time 0. Since only one collapsed state is present, {\it i.e.}, $M=1$, the sMP method indicates that the SCGF $\phi(\lambda)$ of the counting statistics is equal to the largest real root of an algebraic equation~\cite{Liu2022}~\footnote{After publication of our previous paper, we found that Brandes obtained a similar equation in 2008 when he proposed that the waiting time distribution between two consecutive transitions of particles contains complementary information to the conventional full counting statistics in quantum transport; see Eq.~(46) therein~\cite{Brandes2008} (or replace $\lambda$ here by  ${\rm i}\lambda$). Because the key ingredient of the sMPs method is also the waiting time distributions, we believe our method has some connections with Brandes' theory.}
\begin{eqnarray}
	\label{twolevelSCGFsatisfiedequation}
	\hat p_{11}(v)=e^{-\lambda},
\end{eqnarray}
where $\lambda$ $\in(-\infty,+\infty)$ is a real parameter of the cumulant generating function for the counting statistics. In Eq.~(\ref{twolevelSCGFsatisfiedequation}), $\hat p_{11}(v)$ is the Laplace transform of $p_{11}(\tau)$ with parameter $v$:
\begin{eqnarray}
	\hat p_{11}(v)=\int_0^\infty p_{11}(\tau)e^{-\tau v}d\tau.
\end{eqnarray}
The function $p_{11}(\tau)$ is the waiting time distribution in which the system starts from $|1\rangle$, successively evolves and eventually collapses to $|1\rangle$ at time $\tau$. The quantum jump trajectory theory~\cite{Breuer2002} gives its expression:
\begin{eqnarray}
	\label{twolevelsystemp00tau}
p_{11}(\tau)&=&k_{2}\|\sigma_{12}\psi(\tau)\|^2 \nonumber \\
&=&k_{2}\|\sigma_{12}\exp(-{\rm i}\tau\hat{H})|1\rangle\|^2,
\end{eqnarray}
where the non-Hermitian Hamiltonian is~\cite{Breuer2002} 
\begin{eqnarray}
	\hat{H}=H-\frac{\rm i}{2} k_{2}\sigma_{21}\sigma_{12}.  
\end{eqnarray}

For the simple two-level system, the exponential operator in Eq.~(\ref{twolevelsystemp00tau}) has a closed analytical expression. Taking its Laplace transform and substituting the result into Eq.~(\ref{twolevelSCGFsatisfiedequation}), we find that the equation reduces to a cubic polynomial, and the SCGF is exactly solved~\cite{Liu2022,Budini2010}. Even so, the exponential operator formula is too specialized to be available in other quantum systems. Therefore, it would be more desirable if the time waiting distribution could be solved by a general computing scheme. To this end, we write the Laplace transform of the wave function $\psi(\tau)$ as  
\begin{eqnarray}
	\label{wavefunction2level}
\hat\psi(v)=\frac{1}{v+{\rm i}\hat H}|1\rangle=\hat c_1(v)|1\rangle +\hat c_2(v)|2\rangle.
\end{eqnarray}
Given the eigenvalues of the operator ${-\rm i} \hat{H}$ as $h_1$ and $h_2$, the coefficient of the wave function directly involved in Eq.~(\ref{twolevelsystemp00tau}) is  
\begin{eqnarray}
\label{c1twolevelsystem}
\hat c_2(v)={\rm i}\frac{\Omega_2}{2}\frac{1}{(v-h_1)(v-h_2)}. 
\end{eqnarray}
Note that the denominator is also a factorized form of the determinant of the operator $v+i\hat H$; that is,   
\begin{eqnarray}
	\label{twolevelsystemdeterminant}
\det (v+{\rm i}\hat H)=v^2+\frac{k_{2}}{2}v+\frac{\Omega_2^2}{4}.
\end{eqnarray}
Then, the eigenvalues $h_i$ ($i=1,2$) are the two roots of the quadratic polynomial; they are either nonrepeated (unequal) or repeated (equal). Next, a partial fraction expansion is performed on $\hat c_2(v)$ to obtain the Laplace transform $\hat p_{11}(v)$. Because this procedure depends on the roots' properties, we need to discuss them separately. For the case of nonrepeated roots, that is, $k_{2}^2\neq 4\Omega_2^2$, a calculation leads to  
\begin{eqnarray}
	\hat p_{11}(v) 
	&=&\frac{k_{2}\Omega_2^2}{4}\frac{1}{\|h_1-h_2\|^2}\left[\frac{1}{ v-2{\Re }(h_1)} + \frac{1}{ v-2{\Re }(h_2)}-2\Re
	\left(  \frac{1}{v-(h_1+h_2^*)}\right) \right]
	\label{Laplacep00oftwolevelsystemnonrepeatedrootsintermediatestep}\\
		&=&\frac{k_2\Omega_2^2}{2} \frac{1}{(v-2h_1)(v-2h_2)(v-h_1-h_2)},
		\label{Laplacep00oftwolevelsystemnonrepeatedroots}
	\end{eqnarray}
where the unequal roots are $h_{1/2}=-(k_{2}\pm \sqrt{k_{2}^2-4\Omega_2^2})/4$ and the asterisk ($*$) represents complex conjugation. To arrive at Eq.~(\ref{Laplacep00oftwolevelsystemnonrepeatedroots}), we use an intermediate step, Eq.~(\ref{Laplacep00oftwolevelsystemnonrepeatedrootsintermediatestep}), which is also beneficial in analyzing asymptotic behaviors of large deviations. For the case of repeated roots, that is, $h\equiv h_{1/2}=k_2/4$ for $k_2^2=4\Omega_2^2$, we have~\cite{Garrahan2010,Budini2011} 
\begin{eqnarray}
	\label{Laplacep00oftwolevelsystemrepeatedroots}
	\hat p_{11}(v) 
	&=&\frac{k_{2}\Omega_2^2}{2}\frac{1}{(v-2h)^3}.
\end{eqnarray} 
It is worth emphasizing that Eq.~(\ref{Laplacep00oftwolevelsystemrepeatedroots}) is indeed a special case of Eq.~(\ref{Laplacep00oftwolevelsystemnonrepeatedroots}), although the previous intermediate equation is no longer valid.  Substituting  Eq.~(\ref{Laplacep00oftwolevelsystemnonrepeatedroots}) into Eq.~(\ref{twolevelSCGFsatisfiedequation}) and applying Cardano's formula for the cubic polynomial equation, the SCGF of the two-level quantum system is~\cite{Liu2022,Budini2010} 
\begin{eqnarray}
	\label{exactSCGFtwolevelrealroots}
	\phi(\lambda)=-\frac{k_{2}}{2}+\sqrt[3]{\frac{q(\lambda)}{2}+\sqrt{\Delta(\lambda)}}+\sqrt[3]{ \frac{q(\lambda)}{2}-\sqrt{\Delta(\lambda)}},
\end{eqnarray}
where $q(\lambda)=k_{2}\Omega_2^2e^\lambda/2$, $\Delta(\lambda)=q^2(\lambda)/4+p^3/27$ is called the discriminant of the cubic algebraic equation~\cite{Speigel1968}, and $p= (k_{2}^2-4\Omega_2^2)/4$~\footnote{For the case of a negative discriminant, the roots of the cubic algebraic equation have trigonometric forms, which are explicitly real~\cite{Speigel1968}. An advantage of the current formula is its generality for arbitrary $\Delta(\lambda)$}. A brief description of the roots of cubic polynomial equations is given in Appendix A.

Figure~(\ref{fig2}) depicts the SCGF~(\ref{exactSCGFtwolevelrealroots}) under two sets of parameters. These curves appear simple: they are monotonically increasing; when $\lambda$ approaches negative infinity, the data tend toward constants, while when $\lambda$ approaches positive infinity, the data increase exponentially (see inset). Because these characteristics are closely related to the large deviation  properties~\cite{Touchette2008}, it is interesting to find the causes. We use Eq.~(\ref{twolevelSCGFsatisfiedequation}) instead of Eq.~(\ref{exactSCGFtwolevelrealroots}) to conduct this analysis. Although using the latter equation is more direct, exact large derivation formulas are very rare in general quantum systems. First, we note that the denominator of Eq.~(\ref{Laplacep00oftwolevelsystemnonrepeatedroots}) is a cubic polynomial. We immediately have   
\begin{eqnarray}
		\phi(\lambda\rightarrow +\infty)&\sim &	-\frac{k_2}{2}+ \sqrt[3]{\frac{k_{2}\Omega_2^2}{2} }e^{\lambda/3}, 	 
	\label{positivinftySCGF2levelsystem}
\end{eqnarray}
which has nothing to do with the roots' properties and is always valid.  The situation is different for the opposite limit $\lambda\rightarrow-\infty$. For $k_{2}>2\Omega_2$, where the two roots are real and unequal,  Eq.~(\ref{Laplacep00oftwolevelsystemnonrepeatedrootsintermediatestep}) implies that    
\begin{eqnarray}
	\phi(\lambda\rightarrow -\infty)&\sim &
	\label{minusinftySCGF2levelsystem}
	-\frac{1}{2}\left(k_2-\sqrt{k_2^2-4\Omega_2^2}\right)+\frac{k_{2}\Omega_2^2}{|k_{2}^2-4\Omega_2^2|}e^{\lambda}.  
\end{eqnarray}
For $k_{2}<2\Omega_2$, where the two roots are complex conjugates, $h_1=h_2^*$, we have    
\begin{eqnarray}
	\label{positivinftySCGF2levelsystem2}
	\phi(\lambda\rightarrow -\infty)&\sim &
	\label{minusinftySCGF2levelsystem}
	-\frac{k_2}{2}+2\frac{k_{2}\Omega_2^2}{|k_{2}^2-4\Omega_2^2|}e^{\lambda}. 
\end{eqnarray} 
In the second term on the right-hand side of  Eq.~(\ref{positivinftySCGF2levelsystem2}), the first coefficient $2$ is due to $\Re{(h_1)}=\Re{(h_2)}$. A comparison between the exact SCGF~(\ref{exactSCGFtwolevelrealroots}) and the asymptotic formulas is shown in Fig.~(\ref{fig2}). Their agreement is satisfactory~\footnote{We do not analyze the large deviations for the case with $k_2=2\Omega_2$ because the SCGF is exactly equal to the right-hand side of Eq.~(\ref{positivinftySCGF2levelsystem})}. Before leaving the two-level system, we want to point out that the first terms of the right-hand sides of Eqs.~(\ref{minusinftySCGF2levelsystem}) and~(\ref{minusinftySCGF2levelsystem}) can be uniformly represented as $2h_{\max}$, and $h_{\max}$ is the maximum value of the real parts of the eigenvalues of the operator  $-{\rm i}\hat H$.  

\begin{figure}
\includegraphics[width=1\columnwidth]{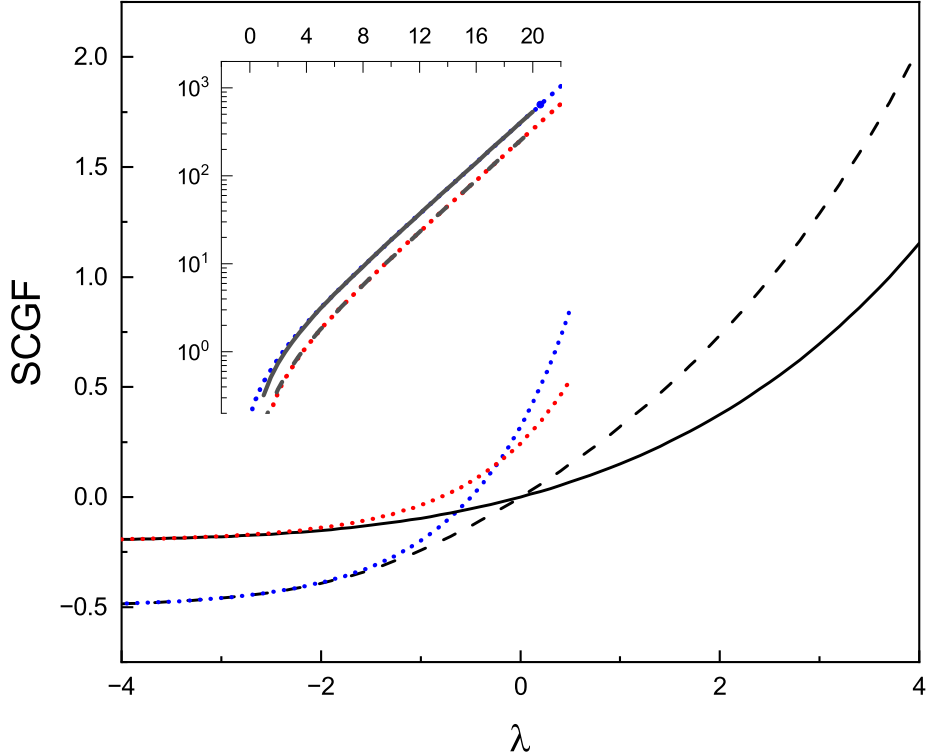}
	\caption{The solid and dashed curves are the data of Eq.~(\ref{exactSCGFtwolevelrealroots}). The parameters are set to $k_{2}=1$, $\Omega_2=0.4$ ($k_2>2\Omega_2$), and $k_{2}=1$, $\Omega_2=0.8$ ($k_2<2\Omega_2$). The dotted curves are the asymptotic behaviors of the SCGF in the limit $\lambda\rightarrow -\infty$. The inset shows the asymptotic behaviors in the opposite limit. A logarithmic scale is used for the SCGFs therein. }
	\label{fig2}
\end{figure}

\section{Three-level system in the $\Lambda$-configuration}
\label{section3}
We apply the previous computing scheme to the three-level quantum system schematically shown in Fig.~(\ref{fig1})(b). The generator of the QME is the same as Eq.~(\ref{QMEtwolevel}) except that the interaction Hamiltonian in this system is 
\begin{eqnarray}
	\label{HamiltonianthreelevelinLambda}
	H=-\frac{\Omega_2}{2}(\sigma_{12} +\sigma_{21})-\frac{\Omega_3}{2}(\sigma_{23} +\sigma_{32}),
\end{eqnarray}
where $\sigma_{23}\equiv|2\rangle \langle 3| $,  $\sigma_{32}\equiv|3\rangle \langle 2|$, and $\Omega_{3}$ is the Rabi frequency of the field driving the $|2\rangle$-$|3\rangle$ transition. The large deviation properties are still determined by Eq.~(\ref{twolevelSCGFsatisfiedequation}). Because this is a three-level system, we set its wave function in the Laplace domain to 
\begin{eqnarray}
	\label{expansionwavefunct3levelsystem}
	\hat{\psi}(v)=\sum_{i=1}^{3}\hat c_i(v)|i\rangle.
\end{eqnarray}
A simple calculation leads us to    
\begin{eqnarray}
	\label{threelevelsystemLambdaLaplacec2}
	\hat c_2(v)&=&{\rm i}\frac{\Omega_2}{2}\frac{1}{(v-h_1)(v-h_2)}. 
\end{eqnarray}
Here, $h_i$, $i=1,2$, are nonzero eigenvalues of the non-Hermitian operator $-{\rm i}\hat H$ of the three-level system. They are also nonzero roots of a cubic polynomial   
\begin{eqnarray}
	\label{threelevelsystemLambdadeterminant}
\det (v+{\rm i}\hat H)=v\left(v^2+\frac{k_{2}}{2}v+\frac{\Omega_2^2+\Omega_3^2}{4}\right), 
\end{eqnarray}
and $h_{1/2}=-(k_{2}\pm \sqrt{k_{2}^2-4(\Omega_2^2+\Omega_3^2)})/4$. When we compare Eqs.~(\ref{threelevelsystemLambdaLaplacec2}) and~(\ref{threelevelsystemLambdadeterminant}) to Eqs.~(\ref{c1twolevelsystem}) and~(\ref{twolevelsystemdeterminant}), we immediately see that the Laplace transform of the time waiting distribution and the following large deviation properties of the three-level quantum system in the $\Lambda$-configuration yield the same conclusions as in the two-level quantum system. We only need to replace $\Omega_2^2$ in the latter by $\Omega_2^2+\Omega_3^2$.  

\section{Three-level system in the V-configuration} 
\label{section4}
\begin{figure}
\includegraphics[width=1\columnwidth]{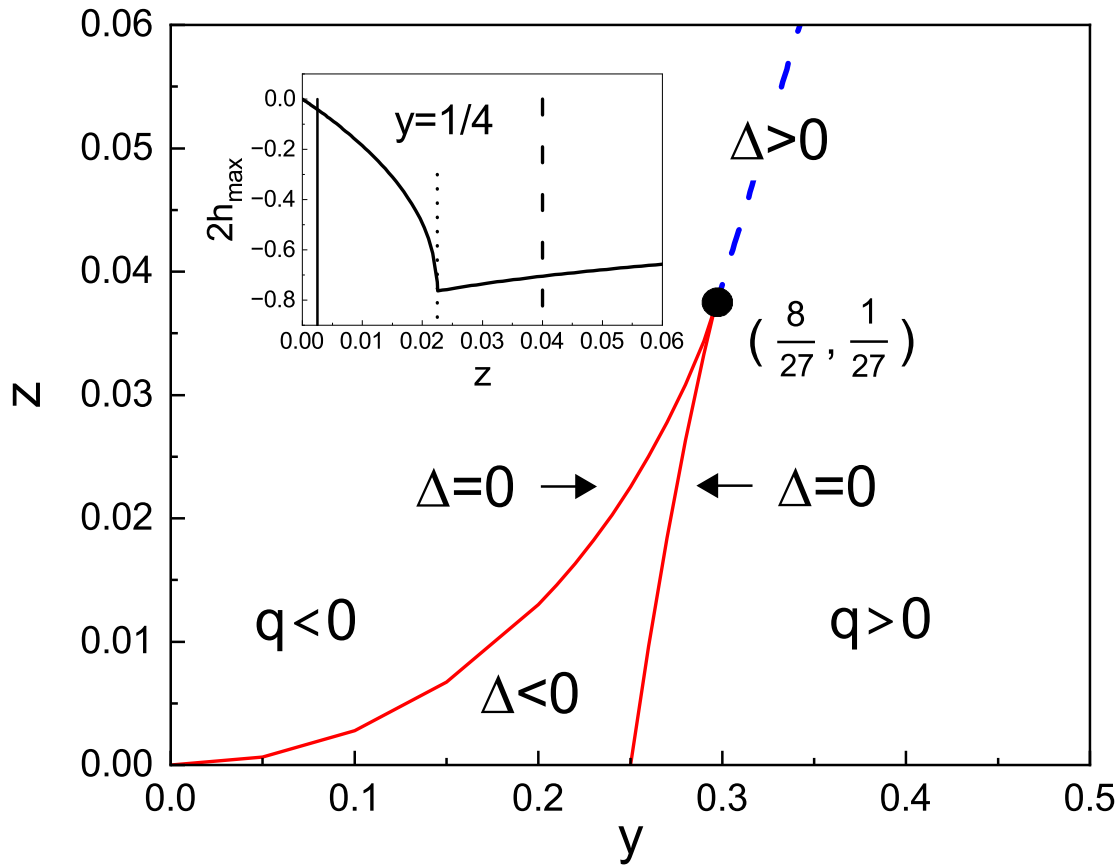}
\caption{Parameter regions where the  discriminant~(\ref{discriminatorcubicequation}) takes negative, positive, and zero values. We see that the region in which $\Delta<0$, or the triangular area enclosed by the two solid red curves ($\Delta=0$) and the $y$-axis, occupies a very small area in the space. The blue dashed line represents the equation $q\propto 9y/2-9z-1=0$. The dark point is special, where $p=q=0$, and $\Delta=0$. The inset depicts $2h_{\max}$ as a function of $z$ with fixed $y=1/4$. The three vertical lines represent the $z$-values of parameter $\Omega_3$ that are used in Fig.~(\ref{fig4}).    }
\label{fig3}
\end{figure}

We now turn to the other three-level quantum system schematically shown in Fig.~(\ref{fig1})(c). The Hamiltonian of the system looks very similar to  Eq.~(\ref{HamiltonianthreelevelinLambda}):
\begin{eqnarray}
	\label{HamiltonianthreelevelinV}
	H=-\frac{\Omega_2}{2}(\sigma_{12} +\sigma_{21})-\frac{\Omega_3}{2}(\sigma_{13} +\sigma_{31}),
\end{eqnarray}
where $\sigma_{13}\equiv|1\rangle\langle 3|$,  $\sigma_{31}\equiv |3\rangle\langle 1|$, and $\Omega_{3}$ is the Rabi frequency of the field driving the $|1\rangle$-$|3\rangle$ transition. We apply the previous computing scheme again. First, the three-level system indicates that Eq.~(\ref{expansionwavefunct3levelsystem}) should be used and the relevant coefficient in the wave function is   
\begin{eqnarray}
	\label{threelevelsystemVLaplacec2}
	\hat c_2(v)&=&{\rm i}\frac{\Omega_2}{2}\frac{v}{(v-h_1)(v-h_2)(v-h_3)}
\end{eqnarray} 
Now, the eigenvalues of the operator $-{\rm i}\hat H$ are $h_i$, $i=1,2,3$. They are also the three roots of the determinant of the operator; that is   
\begin{eqnarray}
	\label{cubiceqthreelevelsystem}
	\det{(v+{\rm i}\hat H)}=v^3+\frac{k_2}{2}v^2+\frac{\Omega_{2}^2+\Omega_{3}^2}{4}v+\frac{k_2}{2}\frac{\Omega_3^2}{4}.   
\end{eqnarray}
Because these three roots distinguish the quantum system in the $V$-configuration from the quantum system in the  $\Lambda$-configuration, we need to analyze them carefully. If we were not concerned about realistic values of the parameters, these roots could be either one real and two complex conjugate roots, or three real roots repeated or nonrepeated. This property depends on whether the discriminant of the cubic polynomial~(\ref{cubiceqthreelevelsystem}),  
\begin{eqnarray}
	\label{discriminatorcubicequation}
	\Delta=\left(\frac{p}{3}\right)^3+\left(\frac{q}{2}\right)^2,
\end{eqnarray}
is positive or nonpositive, where $p=(3b-a^2)/3$,  $q=(9ab-2a^3-27c)/27$,  $a=k_2/2$, $b=(\Omega_{2}^2+\Omega_{3}^2)/4$, and $c=k_2\Omega_3^2/8$. Note that we use the same notations $p$, $q$, and $\Delta$ as in  Eq.~(\ref{exactSCGFtwolevelrealroots}). We do not believe that this will cause confusion. The reason for specifically defining these parameters is given in Appendix A. Interestingly, when we explicitly write the discriminant in terms of the physical parameters, we find that it is simply proportional to   
\begin{eqnarray}
	\label{dimensionlessdiscriminator}
	\left( 3y+3z-1\right)^3+\left(9y/2-9z-1\right)^2,
\end{eqnarray}
where we additionally define two dimensionless quantities,    $y=(\Omega_2/k_2)^2$ and $z=(\Omega_3/k_2)^2$. These two terms in parentheses correspond to dimensionless versions of $(p/3)^3$ and $(q/2)^2$, respectively. In Fig.~(\ref{fig3}), we numerically depict the regions where Eq.~(\ref{discriminatorcubicequation}) takes positive, negative, and zero values.   
Next, a partial fraction expansion is performed on Eq.~(\ref{threelevelsystemVLaplacec2}) to obtain the Laplace transform of the waiting time distribution, $\hat p_{11}(v)$. Because this procedure depends on whether the roots are  repeated or not, we discuss the two cases separately.

\subsection{Nonrepeated roots}
\begin{figure}
\includegraphics[width=1\columnwidth]{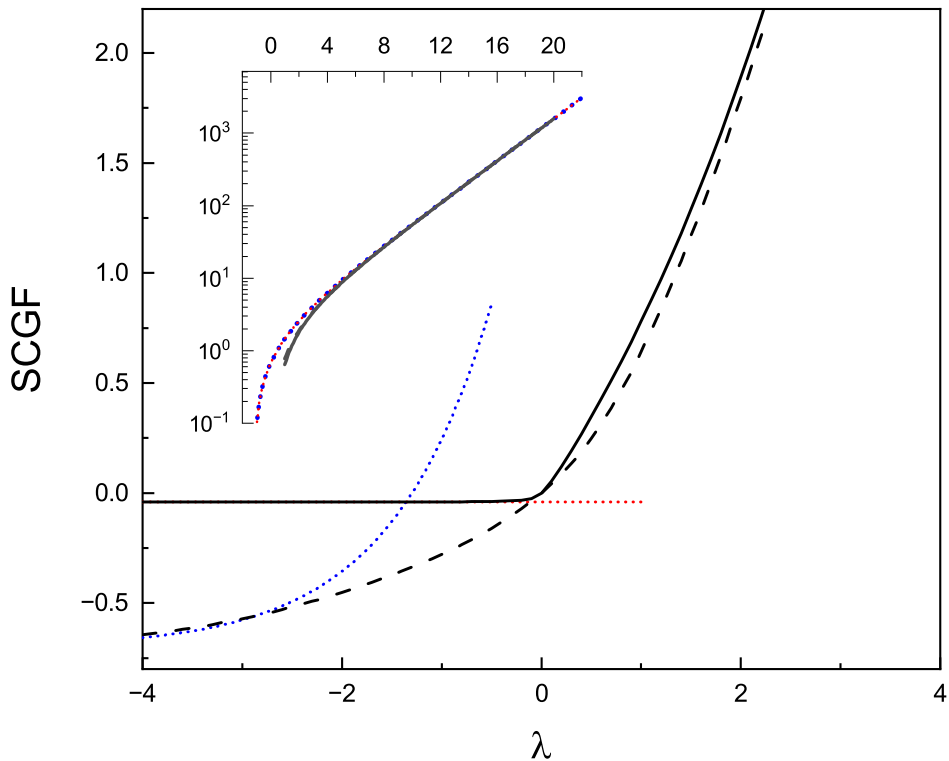}
	\caption{The solid and dashed curves are the SCGFs obtained by numerically solving the sixth-degree polynomial equation. The values of the parameters are $k_2=4$, $\Omega_2=2$, and $\Omega_3=0.2$ for $\Delta<0$ and are $k_1=4$, $\Omega_2=2$, and $\Omega_3=0.8$ for $\Delta>0$ and $q<0$. The large deviation under a specific set of parameters ($\Delta<0$) was discussed in Ref.~\cite{Garrahan2010}, where the TQME method was used. The dotted blue and red curves show the asymptotic behaviors of the SCGFs in the limit $\lambda\rightarrow -\infty$. Note that for $\Delta<0$, we only retain the first constant term and discard the exponential term in Eq.~(\ref{asymptoticSCGFminusinfityDeltanonzero}), since it would induce a dramatic deviation from the data within the parameter range shown in the figure~\footnote{In this situation, we can present an alternative asymptotic formula that is more satisfactory in numerical precision by retaining another term proportional to $1/(v-h_1-h_2)$, where the three roots have the order $h_3<h_2<h_1$. Because the numerical aspects of the large deviations are beyond the scope of this paper, we do not show the results.}. The inset compares the asymptotic behaviors and numerical data in the opposite limit, where a logarithmic scale is also used for the SCGFs.  }
	\label{fig4}
\end{figure}

For this case, after a simple calculation, we obtain  
\begin{eqnarray}
\hat p_{11}(v) &=&\frac{k_2\Omega_2^2}{4}\left[
\sum_{i=1}^3 \|a_i\|^2\frac{1}{v-2\Re (h_i)}+2\sum_{i=1}^3\sum_{j>i}^3 \Re\left(a_ia_j^*\frac{1}{v-(h_i+h_j^*)}\right)\right] 	\label{threelevelsystemLPp10intermediatestep}
\\
&=& \frac{k_2\Omega_2^2}{2} \frac{k_2\Omega_3^2/8+
\prod_{i=1}^{3}(v-h_i)}
 { 	\prod_{i=1}^3\prod_{j\ge i}^3[v-(h_i+h_j)] },
 \label{threelevelsystemLPp10combinedform}
\end{eqnarray}
where $a_{i}=h_i/\prod_{j}'(h_i-h_j)$ are the coefficients of the fractions and the prime symbol means that $i$ is not equal to $j$. For the two-level system, we also use an intermediate step, Eq.~(\ref{threelevelsystemLPp10intermediatestep}). To derive Eq.~(\ref{threelevelsystemLPp10combinedform}), we apply Vieta's formulas for the cubic polynomial~(\ref{cubiceqthreelevelsystem})~\cite{Littlewood1971}. Substituting this result into Eq.~(\ref{twolevelSCGFsatisfiedequation}), we see that the SCGF is the largest real root of a sixth-degree polynomial equation. In general, it is not solvable with  radicals~\cite{Littlewood1971}. Hence, a numerical algorithm for solving roots is needed. Figure~(\ref{fig4}) shows data under two sets of parameters.  

We see that these curves of large deviations are similar to those in the two-level system, which includes approaching exponential functions and some constants in limits of very large $|\lambda|$. If the parameter $v$ is positive and very large, the numerator and denominator of Eq.~(\ref{threelevelsystemLPp10combinedform}) are simply proportional to $v^3$ and $v^6$, respectively. Hence we have 
\begin{eqnarray}
	\label{SCGFthreelevellambdapositiveinfinity}
	\phi(\lambda\rightarrow+\infty)\sim-\frac{k_2}{3}+\sqrt[3]{\frac{k_2\Omega_{2}^2}{2}}e^{\lambda/3}. 
\end{eqnarray}
This explains the former characteristic. 
The latter is implied in  Eq.~(\ref{threelevelsystemLPp10intermediatestep}). For convenience, let us define the maximum value $h_{\max}$ of the nonzero real parts of the three roots. This value is determined by the signs of the discriminant $\Delta$ and the $q$ values. Table I explicitly lists them. We especially specify $h_1$ as real therein; see also   Eq.~(\ref{realrootcubicequation}). After simple calculations, we obtain the asymptotic SCGFs 
\begin{eqnarray}
	\label{asymptoticSCGFminusinfityDeltanonzero}
	\phi(\lambda\rightarrow -\infty)\sim 
	2h_{\max} +\frac{k_{2}\Omega_2^2}{4} C e^{\lambda}. 
\end{eqnarray}
Concrete expressions for the parameter $C$ and their conditions are given in Table I.  
\begin{table}
	\caption{Conditions and parameters of the asymptotic SCGFs $\phi(\lambda\rightarrow -\infty)$ in the three-level quantum system in the $V$-configuration. The first three lines concern the same Eq.~(\ref{asymptoticSCGFminusinfityDeltanonzero}). Note that Eq.~(\ref{asymptoticSCGFminusinfityDelta0q0}) is not included here.}
	\begin{tabular}{p{40pt}p{40pt}p{40pt}p{120pt}p{40 pt}}
		\hline
		\centering$\Delta$ & \centering $q$ & \centering $h_{\max}$ &  \centering $C$ & Eq.  \\
		\hline\hline
	\centering	$ > 0$ & \centering $<0$ & \centering $\Re(h_2)$ & \centering $h_1^2/\|h_2-h_1\|^4$ &| \\
		\centering $ > 0$ & \centering $>0$ & \centering   $h_1$& \centering $\|h_2\|^2/[4\Im^2(h_2)\|h_2-h_1\|^2]$& (\ref{asymptoticSCGFminusinfityDeltanonzero}) \\
		\centering  $ <0$ &  $ $ &  \centering   $h_1$ & \centering   $ {h_1^2 }/
		{[(h_1 -h_2)(h_1-h_3)]^2}$& | \\
		\centering  $=0$&\centering  $<0$ & \centering  $h_2$& \centering  $h_2^{2/3}/(h_1-h_2)^{2/3}$&(\ref{asymptoticSCGFminusinfityDelta0qminus})\\
	\centering  $=0$& \centering  $>0$ & \centering  $h_1$& \centering $h_1^2/(h_2-h_1)^4$&(\ref{asymptoticSCGFminusinfityDeltanonzero})\\
	\hline
	\end{tabular}
\end{table}  

\subsection{Repeated roots}
Unlike in the two-level system, the situation of repeated roots under $\Delta=0$ in the three-level system involves more details. The reason is that the method of making a partial fraction expansion depends on the number of repeated roots. If $p=q=0$, there are three equal real roots. Let the root be $h$. We obtain the following partial fraction expansion: 
\begin{eqnarray}
	\label{threelevelsystemLPp10threerepeatedroots}
	\hat p_{11}(v) &=&\frac{k_2\Omega_2^2}{4}\left[
	 \frac{2!}{(v-2h)^3}+\frac{1}{4}\left(h+\frac{k_2}{2}\right)\frac{3!}{(v-2h)^4}+\left(h+\frac{k_2}{2}\right)^2\frac{4!}{(v-2h)^5}\right]. 
\end{eqnarray}
Note that in this case, $h=-k_2/6$,  $\Omega_{2}=\sqrt{8/27}k_2$, and $\Omega_{3}=\sqrt{1/27}k_2$; see also the coordinates of the dark point in Fig.~(\ref{fig3}). 
On the other hand, if $pq\neq 0$, there are only two equal roots. Let them be $h_2$ and unequal $h_1$. we obtain another partial fraction expansion,  
\begin{eqnarray}
	\label{threelevelsystemLPp10tworepeatedroots}
	\hat p_{11}(v) =\frac{k_2\Omega_2^2}{4}&&\left[\frac{a_1^2}{v-2h_1}+\frac{a_{21}^2}{v-2h_2} + 2 a_1 a_{21} \frac{1}{v-(h_1+h_2)}+ 2 a_1 a_{22}\frac{1}{[v-(h_1+h_2)]^2}+ \nonumber\right. \\
	&& \left. 2 a_{21} a_{22} \frac{1}{(v-2 h_2)^2} + a_{22}^2\frac{2}{(v-2h_2)^3} \right] 
\end{eqnarray}
where $a_1=h_1/(h_2-h_1)^2$, $a_{21}=-h_1/(h_2-h_1)^2$, and $a_{22}=h_2/(h_2-h_1)$. We emphasize that if Eqs.~(\ref{threelevelsystemLPp10threerepeatedroots}) and~(\ref{threelevelsystemLPp10tworepeatedroots}) were written in the form of rational functions, they would become the same Eq.~(\ref{threelevelsystemLPp10combinedform}).  Hence, the asymptotic SCFG $\phi(\lambda\rightarrow +\infty)$ in the three-level system still follows Eq.~(\ref{SCGFthreelevellambdapositiveinfinity}). In the opposite limit, for the case of three equal roots, Eq.~(\ref{threelevelsystemLPp10threerepeatedroots}) leads to 
\begin{eqnarray}
	\label{asymptoticSCGFminusinfityDelta0q0}
	\phi(\lambda\rightarrow -\infty)&\sim & 2h  +k_2  \sqrt[5]{\frac{16}{81}} e^{\lambda/5}.
\end{eqnarray} 
For the case of two equal roots, we separate two further  possibilities. One is that for a parameter $q<0$, the asymptotic SCFG is  
\begin{eqnarray} 
	\label{asymptoticSCGFminusinfityDelta0qminus}
	\phi(\lambda\rightarrow -\infty)&\sim & 2h_{\max}  +\sqrt[3]{\frac{k_2\Omega_2^2}{2}}C e^{\lambda/3},
\end{eqnarray} 
where the explicit $h_{\max}$ and $C$ are given in Table I (the second line from the bottom). The other possibility is that the parameter $q>0$. We find that the asymptotic SCGF $\phi(\lambda\rightarrow -\infty)$ has the same expression as Eq.~(\ref{asymptoticSCGFminusinfityDeltanonzero}), and the parameter $C$ changes accordingly; see the last line in Table I. 
  
\section{Asymptotic large deviation rate functions}    
\label{section5}
The asymptotic SCGFs in the above quantum systems have unified forms in the limit of very large $|\lambda|$: 
\begin{eqnarray}
	\label{unifiedasympototictSCGFs}
	\phi(\lambda)&\sim &
	\begin{cases}
		2h_{\max}+ C_-e^{\lambda/n}, &	\lambda\rightarrow -\infty,\\
		\\
		C_+ e^{\lambda/3},
		&\lambda\rightarrow +\infty,	
	\end{cases}	 
\end{eqnarray}
Here, $h_{\max}$ is the maximum nonzero real part of the eigenvalues of the operator $-{\rm i}\hat H$, the number $n$ takes the value $1,3,5$, and the coefficient  $C_+=\sqrt[3]{k_2\Omega_2^2/2}$~\footnote{When $\lambda$ is positive and very large, we discard different finite constants, $-k_2/2$ in Eq.~(\ref{positivinftySCGF2levelsystem}) and $-k_2/3$ in Eq.~(\ref{SCGFthreelevellambdapositiveinfinity}). }.  Considering that the SCGFs are differential, we can directly apply the Legendre transform on the functions to derive rate functions~\cite{Touchette2008} in two limits:
\begin{eqnarray}
	\label{unifiedasympototicratefunctions}
	I(j)&=&\max_{\lambda}\{j\lambda-\phi(\lambda)\}\nonumber\\
	&\sim &
	\begin{cases}
		-2h_{\max}+nj\ln \left[nj/(C_-)\right]-nj,
		&j\rightarrow 0,	
		\\
		\\
		3j\ln\left(3j/C_+ \right)-3j, &	j\rightarrow +\infty,
	\end{cases}	 
\end{eqnarray}
where $j$ is the rate or the time average of the counting number.

\begin{figure}
\includegraphics[width=1\columnwidth]{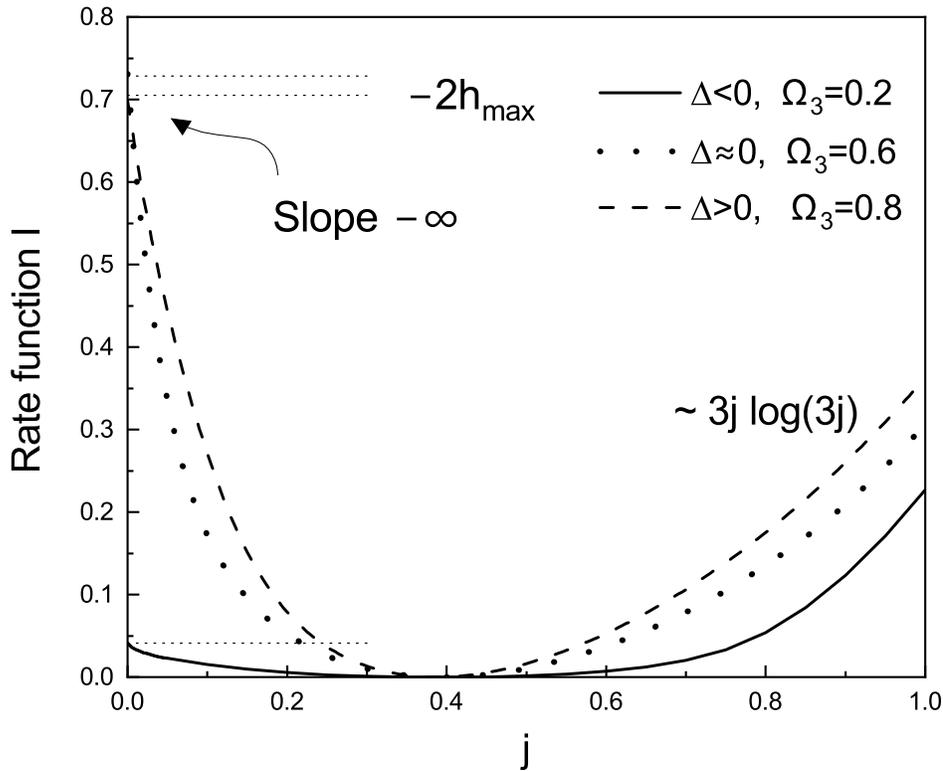}
	\caption{The dashed, dotted, and solid curves are the large deviation rate functions of the three-level quantum system in the $V$-configuration. Except for the Rabi frequency $\Omega_3$, the parameters are the same as those in Fig.~(\ref{fig4}). In particular, we denote the asymptotic behaviors of the rate functions by drawing horizontal dotted lines, marking the slope and formula. Note that the mean rates or the minimum values of these rate functions are almost constant ($\approx 0.4$).  }
	\label{fig5}
\end{figure}

This simple asymptotic rate function provides insights into  the large deviation properties of the counting statistics of these open quantum systems. We see that at a zero rate, the function is exactly equal to $-2h_{\max}$, and its slope is negative infinity. According to the large deviation principle, in the long time limit, the probability distribution of the rate is $p(j)\sim \exp[-tI(j)]$~\cite{Touchette2008}. Because the rate is always nonnegative, when it approaches zero, it is not very surprising that the rate function $I(j)$ tends toward certain constants. However, the physical meanings of these constants and the method of approaching them are nonintuitive. On the other hand, for the three-level system in the $V$-configuration, the combination of Fig.~(\ref{fig3}) and Table I shows that $h_{\max}$ may be a nondifferentiable function on some Rabi frequencies. Indeed, this characteristic arises at the curve made of the upper red solid line and dashed blue line in  Fig.~(\ref{fig3}). As an illustration, the inset therein depicts the function in terms of $z$ by fixing $y=1/4$. We see that when the variable $z$ is larger than a certain value, where the determinant $\Delta$ is zero, $h_{\max}$ is almost fixed at a constant, and when the variable crosses the value from the right and approaches zero, the function rapidly increases to zero. According to  Eq.~(\ref{unifiedasympototicratefunctions}), there is an apparent transition from very improbable to probable in the random event of zero counting. This argument is verified numerically in Fig.~(\ref{fig5}), where the three rate functions of the SCGFs at different $\Omega_3$-values are depicted. 

From the perspective of quantum trajectories, the above observation means that if these counting processes are long but finite in time, at a larger Rabi frequency $\Omega_3$ on the  $|1\rangle-|3\rangle$ transition, wave function collapses occur at a high frequency in almost all trajectories. As $\Omega_3$ decreases and becomes less than a certain value, trajectories with very few collapses appear significantly often. This transition is rapid~\footnote{The data in Fig.~(\ref{fig5}) also indicate that the minimum values of the rate functions are almost constant ($\approx 0.4$) with respect to different $\Omega_3$-values. Hence, if time approaches infinity, quantum trajectories with near-zero rates become be extremely improbable.} In quantum physics, the situation of few collapses indicates that most of the time, the wave functions of the three-level quantum system are in a superposition in which the metastable state $|2\rangle$ is predominant~\cite{Plenio1998}. Therefore, this large deviation shares a common physical foundation with the famous intermittent fluorescence
~\cite{Dehmelt1975,Nagourney1986,Sauter1986,Bergquist1986}. Note that one collapse of the wave function denotes that one photon is emitted.  According to an earlier theory~(\cite{Plenio1998} and references therein), for the three-level quantum system studied in this paper, the parameter condition for exhibiting bright and dark periods is $z\ll y^2$. Obviously, the Rabi frequencies in the triangular area of Fig.~(\ref{fig3}) satisfy this condition. Even so, because the large deviation with zero rate is a fluctuation effect of the stochastic processes, we do not think that  intermittent fluorescence is the cause of the former, or vice versa.

Before closing the main text, we want to mention that the sMP method is also useful in analytically studying SCGFs near $\lambda=0$; see Appendix B. The results therein are intimately related to the full counting statistics of open quantum systems~\cite{Bagrets2003,Brandes2008}. 
  
\section{Conclusion}
\label{section6}
In this paper, we study large deviations of the counting statistics of three open quantum systems that are typical in quantum optics and quantum thermodynamics. The sMP method is used, and a general computing scheme is proposed. We find that in the asymptotic limits, the SCGFs of these systems possess consistent features. Hence, we can unify the large deviation rate functions to simple formulas in zero and large rate limits. An exact connection between the large deviations and the non-Hermitian Hamiltonians of the open quantum systems is naturally revealed by these formulas. In principle, the TQME can be applied to obtain the same results. After all, there is an overlap between these two methods. However, we do not think that this process would be simple. First, one immediately faces an eigenvalue problem for matrices of large size, e.g., a $9\times 9$ matrix for the three-level quantum systems. Even though this eigenvalue problem is successfully transformed into a high-degree polynomial equation, because its coefficients are very complex combinations of microscopic parameters in the QME generator, the connection between the maximum real root of the polynomial equation and underlying quantum physics inevitably becomes ambiguous.

In concretely solving the large deviations of the open quantum systems, we also note that although the sMP method circumvents the process of finding the largest eigenvalue of the generator of a TQME, the method has to perform a Laplace transform on the waiting time distribution and solve the largest real roots of higher-degree polynomial equations.  The three-level quantum system in the $V$-configuration shows that the latter is not easy. Hence, the true benefits of the sMP method are that it has transparent probability interpretations and that it studies specific questions such as asymptotic large deviations in an analytically tractable way rather than in terms of numerical precision or efficiency. 

Thus far, none of our calculations and discussions include the effects of field detuning or the presence of other decay or  excitation channels. Whether these factors alter asymptotically large deviations or make the general computing scheme ineffective is an interesting question that deserves further study. \\

{\noindent Acknowledgments.}
This work was supported by the National Natural Science Foundation of China under Grant Nos. 12075016 and 11575016.\\

\appendix

\section{Three roots of a cubic polynomial}
Given a cubic polynomial with real coefficients on $x$,
\begin{eqnarray}
	x^3+a x^2+ b x+c=0. 
\end{eqnarray}
Its three roots are given by Cardano's formulas~\cite{Speigel1968}: 
\begin{eqnarray}
	x_1&=&-\frac{a}{3}+S+T, 
	\label{realrootcubicequation}\\
	x_2&=&-\frac{a}{3}-\frac{1}{2}(S+T)+{\rm i}\frac{\sqrt{3}}{2}(S-T),\\
	x_3&=&-\frac{a}{3}-\frac{1}{2}(S+T)-{\rm i}\frac{\sqrt{3}}{2}(S-T),
\end{eqnarray}
where 
\begin{eqnarray}
	S&=&\sqrt[3]{q/2+\sqrt{\Delta}},\\
	 T&=&\sqrt[3]{q/2-\sqrt{\Delta}}
\end{eqnarray} 
$p=(3b-a^2)/3$, and $q=(9ab-2a^3-27c)/27$;
$\Delta$ is the discriminant and is defined in Eq.~(\ref{discriminatorcubicequation}). The variables $a,b,c,p,q$ are also used in the main text.

\section{SCGFs of open quantum systems at zero $\lambda$}
Because the SCGF $\phi(\lambda)$ satisfies Eq.~(\ref{twolevelSCGFsatisfiedequation}), we can successively take derivatives of the equation to obtain a sequence of new equations, e.g., 
\begin{eqnarray}
&& \partial_v\hat p_{11}(v)\phi'(\lambda)=-e^{-\lambda},
\label{takederivativehatp11A}\\
&&\partial_v^2\hat p_{11}(v)\left[\phi'(\lambda)\right]^2+\partial_v\hat p_{11}(v)\phi''(\lambda)=e^{-\lambda},
\label{takederivativehatp11B}\\
&&\partial_v^3\hat p_{11}(v)\left[\phi'(\lambda)\right)]^3+3\partial_v^2\hat p_{11}(v)\phi'(\lambda)\phi''(\lambda)+\partial_v\hat p_{11}(v)\phi'''(\lambda)=-e^{-\lambda}, 
\cdots 
\label{takederivativehatp11C} 
\end{eqnarray}
Here, we let $v=\phi(\lambda)$ and exhibit the first three equations. Additionally, the Laplace transform of the waiting time distribution $\hat p_{11}$ can be expanded in terms of $v$:  
\begin{eqnarray}
	\label{TaylorexpansionLaplacTrasfromp11}
	\hat p_{11}(v)=1-\overline {\tau} v +\frac{\overline{\tau^2}}{2!} v^2 - \frac{\overline{\tau^3}}{3!} v^3 +\cdots . 
\end{eqnarray} 
The coefficients $\overline {\tau^n}$, $n=1,2,3,\cdots$, in Eq.~(\ref{TaylorexpansionLaplacTrasfromp11}) are the $n$-th moments of the waiting time; that is, 
\begin{eqnarray}
	\overline{\tau^n}=\int_0^\infty t^n  p_{11}(t)dt=(-1)^n\left.\partial_v^n \hat p_{11}\right|_{v=0}.
\end{eqnarray}
Note that $\hat p_{11}(v=0)=1$ due to the  normalization condition of the distribution. Setting $\lambda=0$ on both sides of Eqs.~(\ref{takederivativehatp11A})-(\ref{takederivativehatp11C}) and using the condition $v=\phi(0)=0$, we derive  
\begin{eqnarray}
\phi'(0)&=&\frac{1}{\overline  \tau}, 
\label{phiprime10}\\
\phi''(0)&=&\frac{1}{{\overline \tau}^3}
		\left(\overline{\tau^2}-\overline \tau^2\right) ,
\label{phiprime20}\\
\phi^{'''}(0)&=&\frac{1}{{\overline \tau}^5}\left[\overline{\tau}^4+3\left(\overline{ \tau^2}\right)^2-3\left(\overline{\tau^2}\right)\overline\tau^2-{\overline\tau}\left(\overline{\tau^3}\right) \right],\cdots 
\end{eqnarray}
In large deviations theory~\cite{Touchette2008}, $\phi'(0)$ denotes the mean counting rate $\overline j$, $\phi''(0)$ represents the fluctuation strength of the counting rate, and $\phi'''(0)$ measures the degree of asymmetry of the counting distribution. Budini obtained Eqs.~(\ref{phiprime10}) and~(\ref{phiprime20}) earlier in another scenario~\cite{Budini2011}.  

It is worth emphasizing that because we have obtained the Laplace transforms of the waiting time distributions, Eqs.~(\ref{Laplacep00oftwolevelsystemnonrepeatedroots}) and~(\ref{threelevelsystemLPp10combinedform}), these waiting time moments can be directly written in terms of the physical parameters of the quantum systems. In the following, we focus on the derivation of the complex three-level system in the $V$-configuration, and the results for the relatively simple  two-level system are given directly. First, we rewrite $\hat p_{11}$ in  Eq.~(\ref{threelevelsystemLPp10combinedform}) in rational function form:  
\begin{eqnarray}
\hat p_{11} &=&\frac{k_2\Omega_2^2}{2}\frac{\sum_{i=0}^3 \alpha_i v^{3-i} }
{ 	\sum_{i=0}^6 \beta_i v^{6-i} },
\label{threelevelsystemLPp10combinedformrationalfunct}
\end{eqnarray}
We set the coefficients $\alpha_0$ and $\beta_0$ to $1$. In the numerator, the other coefficients are 
\begin{eqnarray}
	 \alpha_1&=&k_2/2, \nonumber \\
	 \alpha_2&=&(\Omega_2^2+\Omega_3^2)/4,\\  \alpha_3&=&k_2\Omega_3^2/4,\nonumber 
\end{eqnarray} 
respectively. In the denominator, the coefficients $\beta_i$, $i=1,\cdots,6$, are complex: they are obtained by Vieta's theorem of the sixth-degree polynomial, where the roots of the polynomial equation are   $2h_1,2h_2,2h_3,h_1+h2,h2+h3,h1+h3$; see also the denominator of Eq.~(\ref{threelevelsystemLPp10combinedform}). For instance, the two simplest of these are 
\begin{eqnarray}
	\beta_1&=&-4(h_1+h_2+h_3),
	\label{beta1} \\
	\beta_6&=& 8h_1h_2h_3(h_1+h2)(h2+h3)(h1+h3).
	\label{beta6}
\end{eqnarray} 
Importantly, $\beta_i$, $i=1,\cdots,6$, are symmetric polynomials in $h_1,h_2,h_3$. This is obvious in Eqs.~(\ref{beta1}) and (\ref{beta6}). Hence, according to the fundamental theorem of symmetric polynomials~\cite{Littlewood1971}, we can further write $\beta_i$ in terms of polynomials in the coefficients of Eq. (\ref{cubiceqthreelevelsystem}); that is, $\alpha_1,\alpha_2,\alpha_3/2$, as follows:   
\begin{eqnarray}
	\beta_1&=& 4 \alpha_1,\nonumber \\
	\beta_2&=& 5\alpha_1^2+8\alpha_2,\nonumber \\
	\beta_3&=& 2\alpha_1^3 + 11 \alpha_1 \alpha_2 + \frac{5}{2}\alpha_3,\nonumber \\ 
	\beta_4&=& 6 \alpha_1^2\alpha_2+4\alpha_2^2 +7\alpha_1\alpha_3,\\
	\beta_5&=&4\alpha_1\alpha_2^2+4\alpha_1^2\alpha_3+2\alpha_2\alpha_3,\nonumber \\
	\beta_6&=&4\alpha_1\alpha_2\alpha_3-2\alpha_3^2. \nonumber
\end{eqnarray}
Finally, on the basis of  Eq.~(\ref{threelevelsystemLPp10combinedformrationalfunct}), we obtain the first three waiting time moments expressed in terms of the physical parameters:  
\begin{eqnarray}
	\overline \tau&=&\frac{k_2\Omega_2^2}{2}\frac{\alpha_3\beta_5-\alpha_2\beta_6}{\beta_6^2},\\
	\overline{ \tau^2}&=&{k_2\Omega_2^2}\frac{\beta_6\left(\alpha_1\beta_6-\alpha_2\beta_5\right)+\alpha_3\left(\beta_5^2-\beta_4\beta_6\right)}{\beta_6^3},\\
	\overline{\tau^3}&=&{3k_2\Omega_2^2}\frac{\beta_6\left(\alpha_2\beta_5^2-\alpha_2\beta_4\beta_6-\alpha_1\beta_5\beta_6+\beta_6^2\right)+\alpha_3\left(2\beta_4\beta_5\beta_6-\beta_5^3-\beta_3\beta_6^2\right)}{\beta_6^4},
\end{eqnarray}
We examine these tedious formulas by numerically calculating the quadratic approximation of the rate function, that is,
\begin{eqnarray}
I(j)\approx  \frac{1}{2}I''(\overline{j})(j-\overline{j})^2 
\end{eqnarray}
and $I''(\overline j)=1/\phi''(0)$~\cite{Touchette2008}, and comparing it to the data in Fig.~(\ref{fig5}). We see that the fluctuation strength $\phi''(0)$ increases rapidly in a differentiable way when the Rabi frequency $\Omega_3$ is smaller than a certain value  ($\approx 0.6$) where $\Delta=0$ and further decreases. For the two-level quantum system, the first three waiting time moments are calculated in the same way, and their expressions are as follows: 
\begin{eqnarray}
\overline{\tau}&=&\frac{k_2^2+2\Omega_2^2}{k_2\Omega_2^2} ,\\
\overline{\tau^2}&=&\frac{2k_2^4+2k_2^2\Omega_2^2+8\Omega_{2}^4}{k_2^2\Omega_2^4},\\
\overline{\tau^3}&=& \frac{6k_2^6+72k_2^2\Omega_2^4+48\Omega_{2}^6}{k_2^3\Omega_2^6}.
\end{eqnarray}
Here, we write them directly in terms of the physical parameters $k_2$ and $\Omega_2$.


\end{document}